# Generation of Ensembles of Individually Resolvable Nitrogen Vacancies Using Nanometer-Scale Apertures in Ultrahigh-Aspect Ratio Planar Implantation Masks

Igal Bayn[†‡§], Edward H. Chen[†§], Matthew E. Trusheim[†§], Luozhou Li[†], Tim Schröder[†], Ophir Gaathon[†‡], Ming Lu[∥], Aaron Stein[∥], Mingzhao Liu[∥], Kim Kisslinger[∥], Hannah Clevenson[†], and Dirk Englund[†*]

[†] Department of Electrical Engineering and Computer Science, and Research Lab of Electronics, Massachusetts Institute of Technology, 77 Massachusetts Ave., Building 36-575. Cambridge, MA 02139, USA

[‡] Department of Electrical Engineering, Columbia University, New York, NY 10027, USA

[∥] Center for Functional Nanomaterials, Brookhaven National Laboratory, Upton, NY 11973, USA

**A central challenge in developing magnetically coupled quantum registers in diamond is the fabrication of nitrogen vacancy (NV) centers with localization below ~20 nm to enable fast dipolar interaction compared to the NV decoherence rate. Here, we demonstrate the targeted, high throughput formation of NV centers using masks with a thickness of 270 nm and feature sizes down to ~1 nm. Super-resolution imaging resolves NVs with a full-width maximum distribution of 26±7 nm and a distribution of NV-NV separations of 16±5 nm.**

Among the hundreds of color centers in diamond[1], the NV is the only defect reported which has an electron spin triplet that can be optically initialized[2] and measured[3] using optically detected magnetic resonance (ODMR). The electron spin levels can be manipulated by microwave pulses[4] and exhibit coherence times exceeding milliseconds at room temperature[5,6] and approaching one second at liquid nitrogen temperature[13]. These exceptional properties have enabled demonstrations of qubit gates[7,8], quantum registers[9-11], and NV-photon[12] and NV-NV[13] entanglement. In particular, entanglement between two NV centers coupled via dipolar interactions was recently demonstrated at room temperature[13]. However, to extend this approach to larger numbers of coupled qubits, a technique for fabricating small ensembles of several NVs with separations of 5-20 nm is required, to enable dipolar coupling faster than the NV electron spin decoherence rate [14-21]. These ensembles need to be sufficiently isolated from other NVs and other atomic defects, in particular nitrogen atoms to avoid unnecessary background fluorescence and magnetic noise. One established way to realize such isolated clusters of spins is by nitrogen implantation and subsequent annealing[22]. Recently, spatially selective implantation of NVs has been reported based on nitrogen implantation through an electron-beam patterned resist mask (made of poly methyl methacrylate (PMMA))[14,23], a mica nano-channel hard mask[24], a pierced atomic force microscope (AFM) tip[25], and directly by focused ion beam (FIB)[26]. The smallest reported full-width half-maximum (FWHM) of implanted NV ensembles to date is ~25 nm, achieved by sequential N implantation through a pierced AFM tip; however, individual NV centers were not resolved in this study[25]. By contrast, PMMA masks produced by electron beam lithography (EBL) enable high fabrication rate and implantation with single NV localization, but with a much broader FWHM of ~60-80 nm[6, 20]. Furthermore, to achieve high implantation isolation outside of the defined apertures, a high aspect ratio of the mask is required, as is the case for mica masks[24].

Here, we present an implantation technique based on masks produced from 270nm-thick silicon-on-oxide (SOI) membranes by a combination of EBL lithography and atomic layer deposition (ALD), enabling arbitrarily narrow (measured here below 1 nm) implantation lines with high aspect ratio. This approach combines the low FWHM of the AFM tip implantation with the high throughput of EBL patterning, while achieving the excellent aspect ratio of randomly patterned implantation masks (mica[24]). We employ these masks to fabricate rectangular NV implantation regions in which the smaller dimension has a FWHM as low as 1 nm, reaching a regime where the N distribution is no longer limited by the feature size of the mask but by the fundamental process of implanted nitrogen scattering in the diamond lattice. This approach opens the door to scalable fabrication of isolated spin ensembles for quantum information processing and quantum transport[19] measurements.

State-of-the-art EBL is capable of writing features as small as ~5 nm directly into the e-beam resist[27, 28]. However, such patterning requires thin resist (10 nm[27, 28]), which does not provide sufficient isolation between the target zones inside the apertures and the masked areas. For instance, an implantation depth of 10 nm into the diamond would require a PMMA thickness of

~75 nm to achieve an isolation factor of 100, the lower limit for adequate background reduction. The aspect ratio of a resist can in principle be increased by reactive ion etching (RIE) pattern transfer into a second mask, but material removal with RIE becomes extremely inefficient for hole sizes on the nanometer scale and depths of tens to hundreds of nm (see Supplementary) due to lateral etching. Another strategy for patterning small apertures employs electron beam ablation of thin membranes of silicon nitride[29, 30]. For this method, the depth of such apertures is limited to 10-15 nm, and the time-intensive nature of this approach makes it unsuitable for patterning large areas. As present fabrication methods face similar limitations as those discussed above, it is evident that presently available scanning beam patterning and dry etching techniques do not produce small enough apertures with sufficient large aspect ratio. To overcome this limitation, we have developed a technique in which apertures patterned into a Si membrane by standard EBL and RIE (Fig. 1a) are subsequently narrowed by conformal ALD. In this way, several 10 nm wide lines — well within standard silicon patterning processes — were narrowed, with atomic layer precision control, to the sub-nanometer scale by ALD of alumina ($Al_2O_3$). These silicon masks were subsequently released from the Si substrate by hydrofluoric acid (HF) etching of the 3μm $SiO_2$ layer in the SOI chip (SOITEC), and then mechanically transferred onto high-purity diamond substrates (Element6; nitrogen concentration below 10 ppb) either before or after ALD deposition. Because the mask is fabricated separately from the diamond, we avoid possible degradation of the diamond substrate surface during the processing steps. After the transfer of these Si-$Al_2O_3$ masks onto the diamond, nitrogen was implanted over a range of doses (2-4×10$^{13}$ ions/cm$^2$) and energies (6-20 keV, Innovion). After mechanically removing the mask, the diamond was annealed at 850°C to form NV centers.

Figure 1a,b compares two processing sequences (denoted A and B). For each sample, a Si mask with implantation apertures of various nominal lengths (L=200-1000nm, in increments of 100 nm) and widths ($W_{Si}$=45-100 nm, in increments of 1-2 nm) underwent ALD before (Sample A) or after (Sample B) mask transfer onto diamond (see Methods). Both approaches have different merits. By first transferring the Si membrane and then depositing $Al_2O_3$ (Sample B), a thin layer of alumina was also deposited on the diamond surface; this can prevent ion channeling effects[14], but also requires higher implantation energy to reach the same depth as in Sample A, which in turns leads to additional implantation straggle. A benefit of ALD deposition before mask transfer (Sample A) is that the same mask can be re-used multiple times. Fig. 1b shows nitrogen distribution profiles simulated by Stopping Range of Ions in Matter (SRIM)[31] software for both samples. These calculations assume implantation energies of 6keV and 20keV for Samples A and B, respectively, and result in a theoretical lateral straggle of 3.1 nm for Sample A and 11 nm for Sample B.

The mask for Sample B was transferred directly from the SOI wafer onto diamond by wet release in hydrofluoric acid (See Methods). This allows arbitrarily large (mm-scale) membranes to be transferred with high yield. Such a large mask allows for simultaneous N implantation into millions of target regions, increasing the probability of creating addressable and dipole-coupled NV spin ensembles and mitigating the impact of local mask processing or transfer induced imperfections. By contrast, ALD before mask transfer (Sample A) requires the membranes to be first suspended on the Si wafer after undercut. To avoid membrane bowing and adhesion to the Si substrate during the wet HF undercut step, we found that membranes should have side lengths below 100 μm. Critical point drying could extend the size of the membranes used in Sample A.

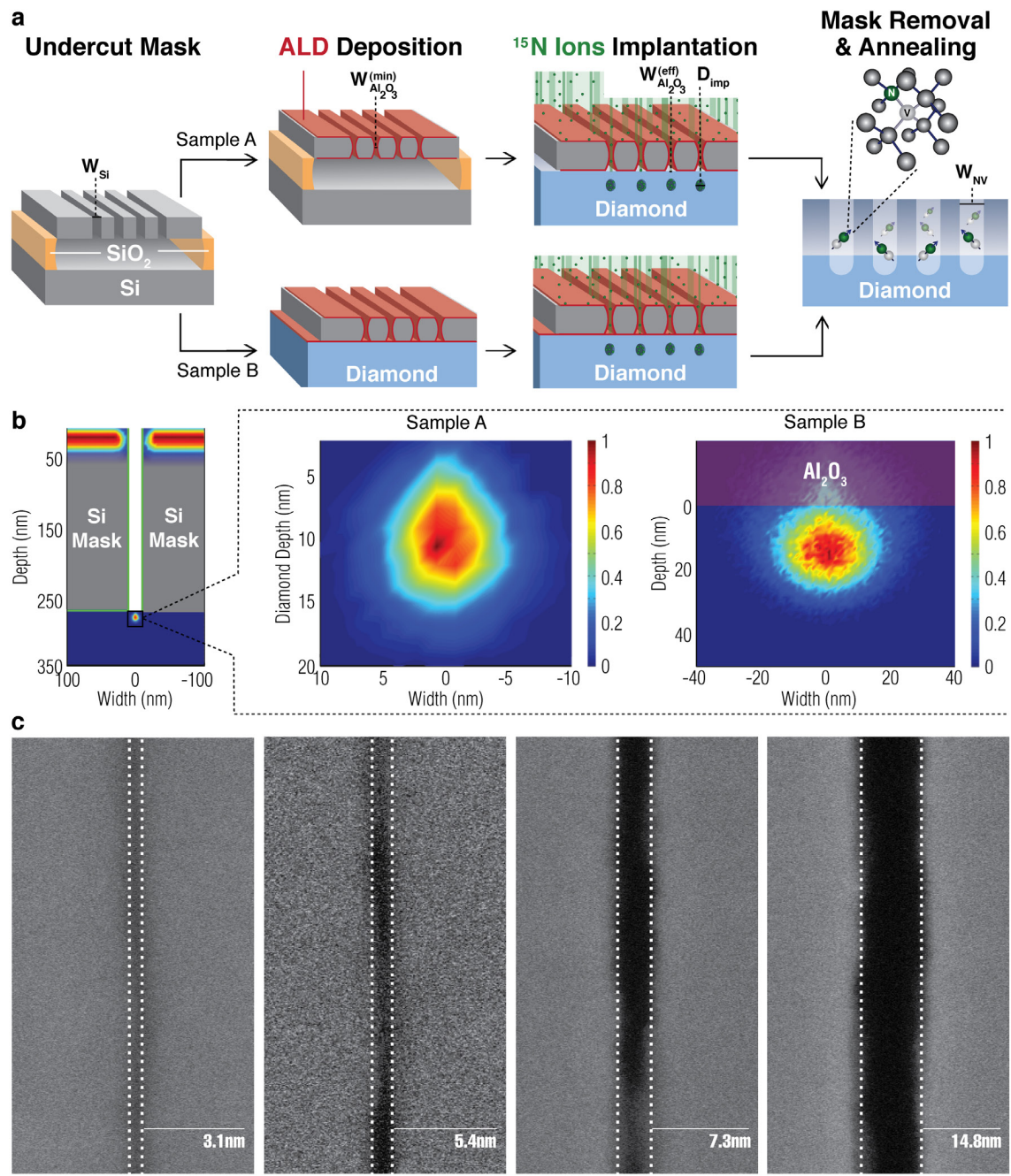

**Figure 1**. **Implantation through silicon hard mask. a.** Fabrication scheme from silicon mask undercut to implantation for Samples A and B. **b**. SRIM simulation of implanted nitrogen. Vertical cross-section of the mask and implantation region (left); close-up of nitrogen post-implantation ion on Sample A (center, 6keV) and Sample B (right, 20keV) with lateral straggle of 3.1 and 11 nm, respectively. Color bar: normalized nitrogen density. **c.** Top-down SEM images of mask following ALD, showing a minimum mask width of 3.1 nm.

Characterization of the membranes using scanning electron microscope (SEM) imaging revealed mask lines with a minimal width of $W_{Si}$ = 25 nm following pattern transfer into a Si membrane with a thickness of 270 nm (see Supplementary). To perform this transfer, cryogenic reactive ion etching was used. The high aspect ratio of the line was achieved via intentional over-etching with a ~3 times increase in the etching time compared to the etching time for larger apertures ($W_{Si}$ > 150 nm). This width, while demonstrating the limitation of silicon-only processing, is already comparable to the AFM tip implantation resolution[25] and is more than two times smaller than that in PMMA masks[14, 23]. Following conformal $Al_2O_3$ ALD, we observed gaps as narrow as 3.1 nm from the top surface on both samples (Figure 1c).

As top-down SEM does not reveal the aperture profile, we examined cross-sections of a Si membrane mask for fabrication process A via focused ion beam (FIB) sectioning. To prevent re-deposition of Si, a 200 nm thick layer of platinum (Pt) was deposited. The high-magnification SEM images in Fig. 2a,b confirm that the gaps were open throughout, with a width down to ~10 nm. We used transmission electron microscopy (TEM) to image with a resolution below 10 nm (see Methods). The TEM images in Fig. 2c-e show the profiles of $W_{Si}$=[45nm, 47nm, 50 nm] after ALD of $Al_2O_3$. The aperture width varies with depth into the Si mask, reaching its narrowest "waist" near the center and widening towards the top and bottom surfaces of the membrane. Fig. 2c shows that apertures can reach waist widths of only $W_{Al_2O_3}^{(min)}$ = 0.9±0.3 nm. This ultra-narrow width was produced by deposition of 22.5 nm thick $Al_2O_3$ on a gap of width $W_{Si}$=45±1 nm. Figure 2d and e show TEMs of apertures with initial widths of $W_{Si}$ =[47±1 nm, 50±1 nm], resulting in gaps of $W_{Al_2O_3}^{(min)}$=[2.2±0.2 nm, 4.4±0.4 nm] (Figure 2d,e). Remarkably, because of the atomic control of the ALD thickness, these $W_{Al_2O_3}^{(min)}$ values are wider than for the $W_{Si}$=45 nm masks by almost precisely the difference in starting masks widths, [2±1 nm, 5±1 nm]= [47±1 nm, 50±1 nm]-45 nm, as one may expect for an ideal deposition process.

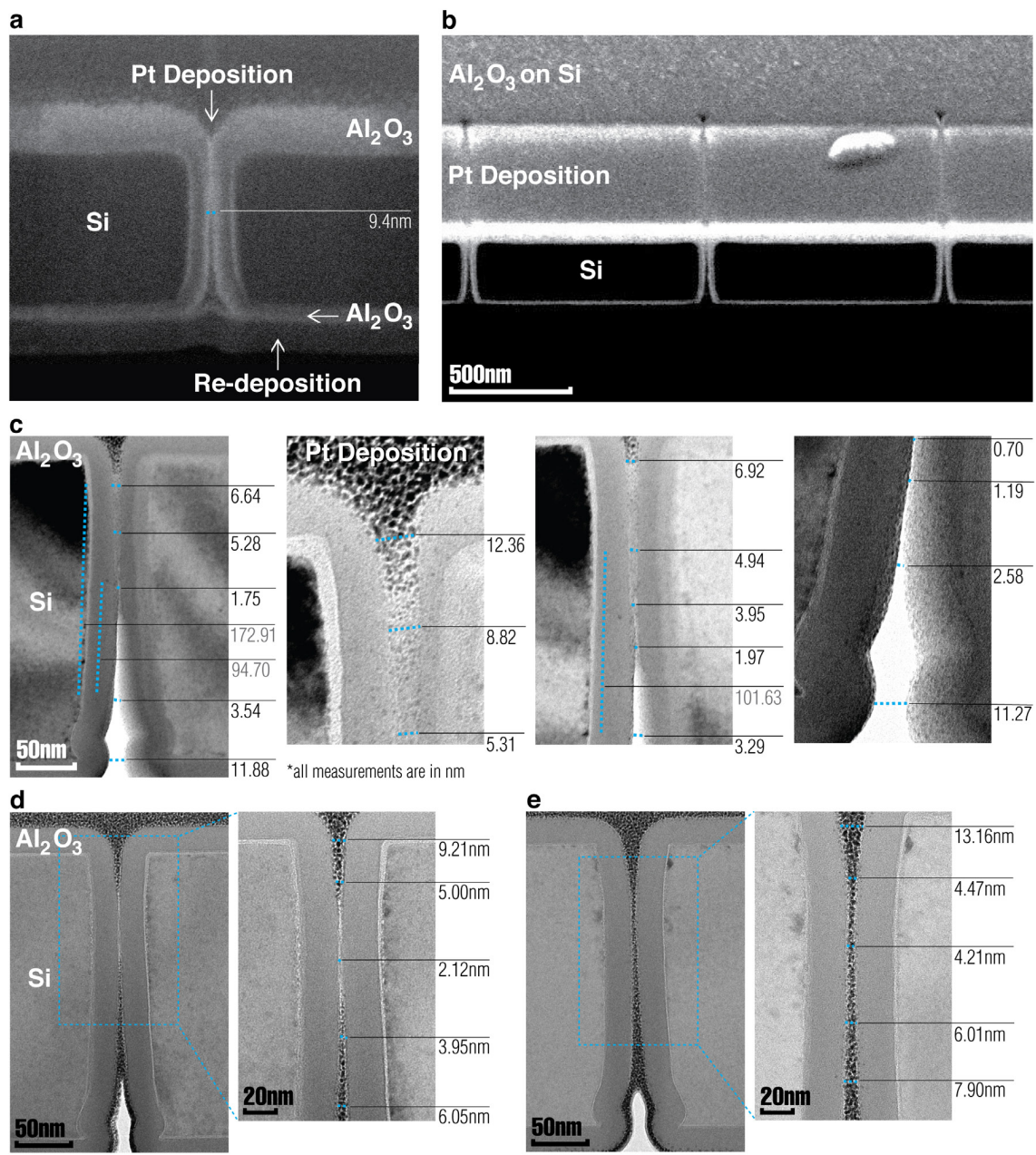

**Figure 2. Vertical profile of ALD coated sub-15 nm mask (Sample A). a-b.** FIB cross-section with Pt layer on top. **a.** 10 nm wide line **b.** 15 nm wide lines. **c-e.** TEM analysis of ALD coated partially undercut masks. **c** $W_{Si}$=45 nm lines, corresponding to $W_{Al_2O_3}^{(min)}$=0.7-1.2 nm. From left to right: TEM sample aperture, top line profile, middle line, bottom. **d, e.** TEM analysis of implantation line with $W_{Si}$=47 and 50 nm, corresponding to $W_{Al_2O_3}^{(min)}$=2 and 4 nm respectively.

After these masks were mechanically transferred onto the target diamond (Fig 3a,b), we performed nitrogen implantation (Innovion) with the parameters shown in Figure 3b. We parameterize the expected nitrogen distribution after implantation by a characteristic length $D_{imp}^{(est)}$, the Gaussian FWHM corresponding to the expected implantation spatial variance: $D_{imp}^{(est)} = W_{Al_2O_3}^{(min)} + A\sqrt{\sigma_{mask}^2 + \sigma_{straggle}^2}$ where the factor of $A = 2\sqrt{log2} \approx 2.35$ converts a standard deviation to full-width half-maximum. The standard deviation $\sigma_{mask}$ is calculated by approximating the mask as a zero-mean uniform distribution with width $W_{Al_2O_3}^{(min)}$, while $\sigma_{straggle}$ is the lateral implantation straggle calculated by SRIM. For our experimental parameters on Sample A (Figure 3c), we expect a nitrogen distribution $D_{imp}^{(est)} \sim 9.1$ nm. Sample B had an increased implantation straggle due to the increased implantation energy, resulting in $D_{imp}^{(est)} \sim 26.3$ nm. In both samples A and B the NV distribution is dominated by straggle rather than the mask dimension, as $\sigma_{straggle} \gg \sigma_{mask}$. Following the implantation, the $Al_2O_3$ coating of Sample B was removed using HF and the masks from both samples were mechanically removed. Subsequently, all samples were annealed at 850°C to create NVs. Fig. 3a,b show fluorescence microscopy images of the resulting NV ensembles for samples A and B, respectively.

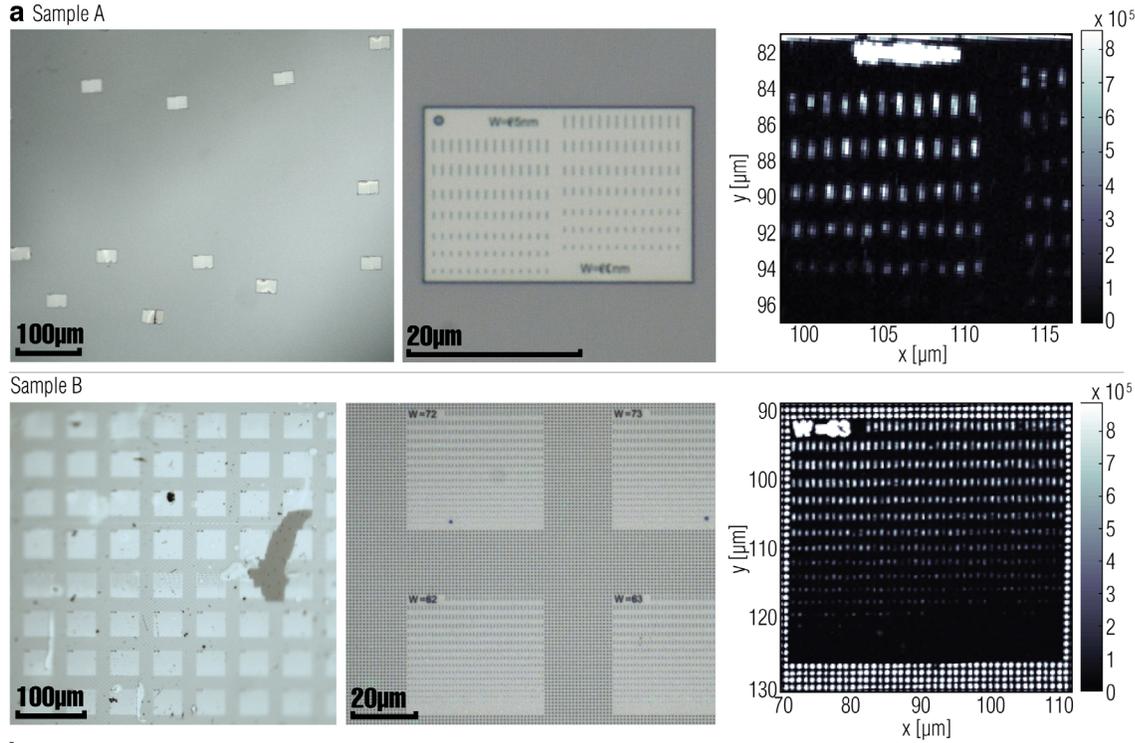

**Figure 3. Sample parameter space summary. a.** Microscope images of Si masks and confocal scans of NV ensembles pre- and post implantation, respectively. Sample A (top) and Sample B (bottom). **b.** Summary of mask and implantation parameters with minimal line width defined by $W_{Al_2O_3}^{(eff)}$, initial silicon width $W_{Si}$, and the corresponding theoretical limit to the implantation distribution $D_{imp}^{(est)}$.

We performed wide-field deterministic emitter switch microscopy (DESM) to image NVs produced by these implantation masks. DESM was used to localize NVs by modulating the spin-dependent fluorescence of NVs in the four crystallographic orientations of single-crystal diamond[18]. An external, locally homogenous magnetic field was used to impart a unique Zeeman shift for each NV orientation, so that NVs in different directions could be individually modulated. Since NVs with identical orientations are indistinguishable in this method, only sites with four or fewer NVs — each with different orientations in the crystal lattice — are spatially resolvable. Therefore, our statistics were assembled only from sites with single NVs in any given crystal orientation. Such sites were determined by the background-free brightness of the spot, normalized to the brightness of single-NV sites. This single-NV-brightness was separately measured from spots in which the second order correlation function at zero time delay, $g^{(2)}(0)$, was less than 0.5 as determined by confocal microscopy with two avalanche photodiodes in a Hanbury-Brown-Twiss configuration. Fig 4a shows an exemplary implantation spot on Sample A that contains a single NV, together with the corresponding ODMR spectrum. In Figure 4b, an ensemble containing three NVs from Sample B is imaged as an example, with the closest NV pair having a separation of 16±5 nm. Wide-field DESM enabled us to image distances as small as 8±3 nm (see Supplementary).

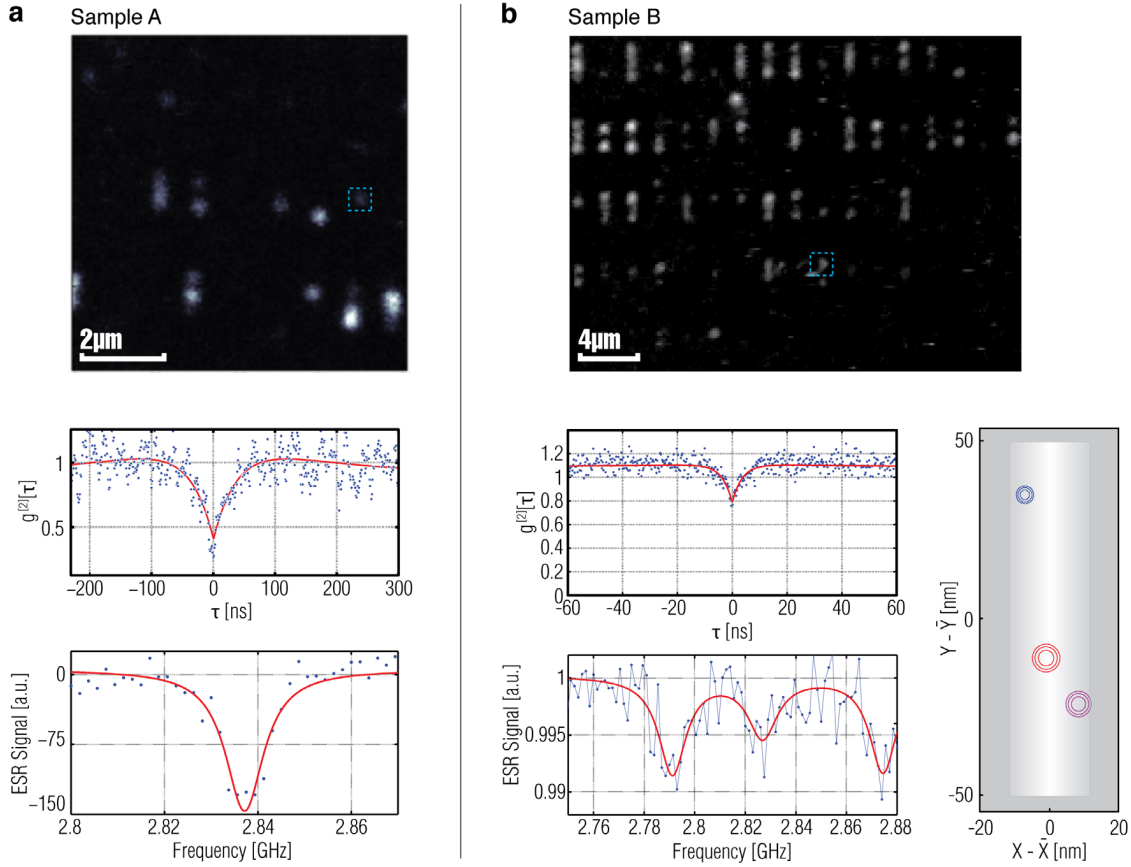

**Figure 4. Analysis of implantation lines. a.** Confocal scans of various length ($W_{Si}$=55 nm, $W_{Al_2O_3}^{(min)}$ = 10 nm) lines of Sample A are shown. Single NV occupied lines are resolved, one is marked with a square dotted line. Its $g^{(2)}$ (middle) and ESR (bottom) are presented. **b.** Confocal scans of various length ($W_{Si}$=54 nm, $W_{Al_2O_3}^{(min)}$< 1 nm) on Sample B are shown. An ensemble of three NVs (L=100 nm) marked with blue dotted square is resolved. Its $g^{(2)}$, ESR and super-resolved NV distribution are shown, with d(NV2,NV3)= 16±5 nm, d(NV1,NV3)= 46.5±5 nm.

We evaluated the localization of the NV ensembles by statistical analysis on multiple spots for which we verified that only single NVs exist in each crystal direction, but with at least two NVs overall. Super-resolution imaging was used to localize NVs in such spots with respect to the estimated lithographically defined center of such implantation holes. For Sample A, where the narrowest lines were populated only at the most by a single NV, this lithographic center line was estimated by a linear fit across multiple ensembles (see Supplementary). The resulting distribution of NVs in the transverse direction relative to the lithographic center ($X_{i,j} - \bar{X}_j$ for all NVs *i* within ensemble *j*) is shown in Figures 5a,b. The standard deviation of this distribution allows for calculation of the ensemble localization in the confined direction, given as a Gaussian FWHM: $D_{NVA}^{(meas)}$=26±3.75 nm for Sample A, and $D_{NVB}^{(meas)}$ = 26±2.4 nm for Sample B. These results agree with predicted distributions given the measured mask widths: $D_{impA}^{(est)}$ = 19.1 nm for sample A and $D_{impB}^{(est)}$ = 26.3 nm for Sample B. Although initial estimation predicted that NV localization on Sample A could be as low as 9 nm for $W_{Al_2O_3}^{(eff)}$ = 1 nm lines, only the lines with $W_{Al_2O_3}^{(eff)} \geq 11$ nm — corresponding to $D_{imp}^{(est)} \geq 19$ nm — had a sufficient probability of containing any NVs to allow for statistically meaningful analysis. To achieve more localized NV ensembles, a higher implantation dosage, improved N to NV conversion yield, or a larger number of implantation holes would be needed. The lateral and axial positions of NVs in Sample B compared to the mask center are shown in Figure 5b. The distribution of NV positions along the lateral axis matches closely the expected nitrogen distribution through an infinitesimally narrow mask, as calculated by SRIM (Figure 5c). The ensembles therefore reach the spatial localization limit set by the implantation straggle rather than any limit set by aperture size. To the best of our knowledge, this is the narrowest NV ensemble localization reported.

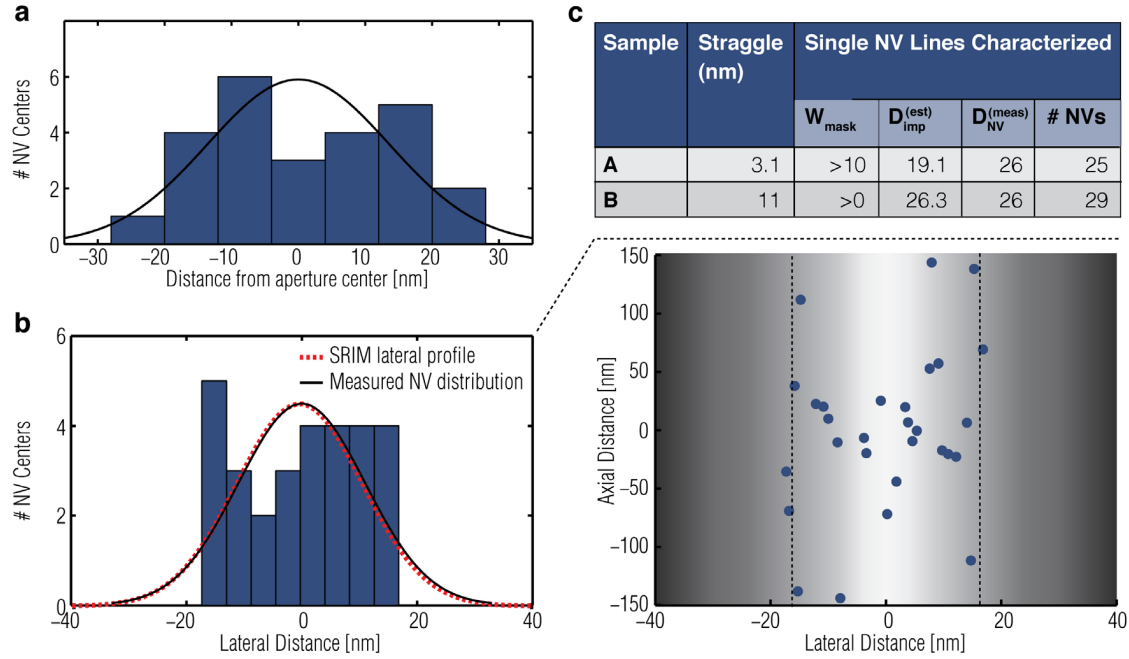

**Figure 5. Line width analysis. a.** Single NV distribution histogram for W=55 nm, $W_{Al_2O_3}^{(min)} \sim 11$ nm L = 500 nm lines (Sample A) as a function of its distance from the aperture center shows FWHM =26±2.4 nm. **b.** Sample B distribution of $W_{NV}$ in NV regions with $W_{Al_2O_3}^{(min)} <$ 1 nm, FWHM = 25.8±3.5 nm. Right: spatial map of measured NV positions relative to ensemble center, individual regions are mapped on top of each other. **c.** Table summarizing mask width, estimated WN distribution and measured NV distribution.

The sub-1 nm mask width demonstrated here approaches the limit corresponding to the thickness of a single alumina deposition cycle. Despite having a sub-nanometer-scale, the lateral straggle increased the estimated NV-localization limit to 7 nm. For quantum information (QI) applications, it is conceivable to achieve improved control over NV spacing along the aperture line by modifying the pattern with a line of connected or partially overlapping, small holes. In this way, after alumina deposition, only circular apertures at given pitches remain, and lateral localization could likely be reduced to 5-10 nm. Since the vertical implantation straggle varies from 3.1 to 11 nm, for the given implantation energies, the NVs would be localized within ~5-11 nm in all directions. Realization of this scheme would likely require a decrease in the thickness of the resist and the membrane, which would also require a decrease in implantation energy.

Because of the stochastic implantation process (see Supplementary), controlling the NV configurations below 10 nm or so is not feasible; for this reason, it is important to produce large number of arrays, as demonstrated in Sample B, and selecting the successful ones. The feasibility of this approach is demonstrated in Supplementary for circular apertures. The NV-NV distances obtained by this method allow for the production of coupled spin systems. For NV-NV separations of 20 nm or less, dipolar coupling strengths on the order of tens of kilohertz are expected. This coupling would allow the entanglement of nearby NVs, with long spin coherence times, which have been shown in shallow-implanted bulk diamond[32, 33] as well as nanodiamond[34].

In this work we have presented a mask implantation technique with minimum dimensions down to ~1 nm. The mask has a thickness in excess of 270 nm and therefore enables ion implantation with high isolation. We demonstrated $^{15}$N implantation at energies of 6 and 20 keV, producing localization of the resulting NVs with a FWHM down to 26±2.4 nm. Such narrow localization enables QI applications based on magnetically coupled NV spin systems. The membranes, which were produced by a combination of EBL-RIE processing and conformal ALD deposition, could be adjusted to match a wide range of implantation energies. In addition, this mask fabrication process is also suitable for dry etching and deposition processes on a wide range of materials, including non-flat substrates. The straggle-limited N implantation and NV formation demonstrated here represents an important step for the precise fabrication of coupled spin systems, with applications in quantum transport and quantum simulation[6, 13, 19], room-temperature quantum computing with small spin ensembles[16], quantum registers[11], and spintronic devices[35].

Methods

Fabrication

270 nm thick SOI with 3μm buried oxide layer was used to produce the hard masks. Upon the SOI surface undiluted ZEP 520A e-beam resist was spun at 4 Krpm and pre-baked at 180°C for 3 min. The mask patterns were written by JEOL JBX-6300FS at 100 kV with beam current of 250 pA at high magnification mode (Mode 6) with a dosage of 600 μC/cm$^2$. Post writing, the resist was developed in cooled Hexyl Acetate at temperature of -25°C for 95 s and cleaned in isopropanol for additional 95 s. Post development, the SOI hard mask are dry etched by Oxford Instruments Plasmalab 100 using mixture of $SF_6:O_2$=40:18 at 100°C with ICP power of 800 W and RF power of 15 W for 30 s. After etching the resist was removed by Remover PG at 90°C. Then, the sample was undercut in 49% concentrated hydro-fluoric acid (HF) for 4 min. At this stage the freestanding 30 μm×20 μm hard masks are attached to the SOI substrate with four 350 nm-wide silicon bridges at the mask corners.

Then, in Sample A, the uniform and conformal coating of alumina to the high aspect ratio holes was applied by an ALD in exposure mode. In this way, the substrate is alternatively exposed to precursor vapors of trimethylaluminum and water for an extended period (~ 5 s) to provide enough diffusion time for the precursors to reach inside the narrow holes. The deposition is performed on CambridgeNanoTech Savanah S100. In this sample the mask lines with a width varying from 30-100 nm with a width step of 2-3 nm were covered by 175 deposition cycles. Then, a thin tungsten probe covered with adhesive micro-spheres of Polydimethylsiloxane (PDMS) and mounted on a micromanipulator was used to detach the masks from the SOI (by breaking the bridges) and to transfer them onto the diamond surface. The diamonds used here are ultra-pure diamond from Element Six with nitrogen concentration lower than 5 ppb. Prior to the transfer, the diamonds were cleaned in piranha solution $H_2SO_4/ H_2O_2$ (3:1).

For sample B the millimeter mask was undercut in 49% HF. Then the HF concentration was diluted and the mask was transferred first onto a Teflon, and then onto the diamond surface. Then, 225 cycles of alumina were deposited in the expose mode while the mask was sitting on the diamond.

The samples were implanted with nitrogen ($N^{15}$) at 6 KeV (Sample A) and at 20KeV (Sample B). The dosages were as described in the paragraph above. Post implantation, sample B was exposed to 49% HF for 5 min to remove alumina.

Then the masks were mechanically removed from all samples and they were annealed at 850°C in vacuum. As the last step the diamonds were cleaned in boiling $HClO_4/HNO_3/H_2SO_4$ (1:1:1).

**Cross-section preparation:** Fully under-cut masks were covered with ~100 nm of Pt to prevent milling re-deposition during cross-section. This was done in situ deposition at Helios NanoLab dual SEM/FIB. Then, vertical cross-sections were milled by FIB in the center of 10 and 15 nm wide lines with 48 pA current at 30 kV voltage.

**TEM sample preparation:** After EBL the masks were dry-etched and exposed to 49% HF for 5 s. This produced the required undercut to enable conformal alumina deposition with ALD in the way similar to our free standing membrane masks, but leaving enough $SiO_2$ substrate beneath the lines to allow for TEM sample preparation and mounting on the grid. The partially undercut samples were covered with 175 cycles of alumina in the expose mode. The samples were thinned to ~50 nm by FIB in the middle of a 1.5 μm long line region. The final ion milling was performed at 5 keV to reduce beam damage to the sample. The thinned lines were analyzed with JEOL JEM-1400 LaB6 120 KeV.

## ASSOCIATED CONTENT

**Supporting Information**. Supplementary is included. This material is available free of charge via the Internet at http://pubs.acs.org.

## AUTHOR INFORMATION

### Corresponding Author

*Correspondence and requests for materials should be addressed to D. E. (email: englund@mit.edu)*

### Author Contributions

The manuscript was written through contributions of all authors. All authors have given approval to the final version of the manuscript. § These authors contributed equally to this work.

**Competing Interests** The authors declare that they have no competing financial interests.

### Funding Sources

Financial support was provided in part by the W.M. Keck Foundation, GTech MURI, and AFOSR PECASE. E.H.C. and H.C. were supported by the NASA Office of the Chief Technologist's Space Technology Research Fellowship. M.T. was funded by the NSF IGERT program Interdisciplinary Quantum Information Science and Engineering. T.S. acknowledges support by the Alexander von Humboldt foundation. This Research was carried out in part at the Center for Functional Nanomaterials, Brookhaven National Laboratory, which is supported by the U.S. Department of Energy, Office of Basic Energy Sciences, under Contract No. DE-AC02-98CH10886.

ACKNOWLEDGMENT

Financial support was provided in part by the W.M. Keck Foundation, GTech MURI, and AFOSR PECASE. E.H.C. and H.C. were supported by the NASA Office of the Chief Technologist's Space Technology Research Fellowship. M.T. was funded by the NSF IGERT program Interdisciplinary Quantum Information Science and Engineering. T.S. acknowledges support by the Alexander von Humboldt foundation. The authors would like to thank Alexey Tkachenko for insightful discussions, Xinwen Yao and Nathalie de Leon for help in diamond preparation. This Research was carried out in part at the Center for Functional Nanomaterials, Brookhaven National Laboratory, which is supported by the U.S. Department of Energy, Office of Basic Energy Sciences, under Contract No. DE-AC02-98CH10886.

# Supplementary Information

# Generation of Ensembles of Individually Resolvable Nitrogen Vacancies Using Nanometer-Scale Apertures in Ultrahigh-Aspect Ratio Planar Implantation Masks

Igal Bayn[†‡§], Edward H. Chen[†§], Matthew E. Trusheim[†§], Luozhou Li[†], Tim Schröder[†], Ophir Gaathon[†‡], Ming Lu[∥], Aaron Stein[∥], Mingzhao Liu[∥], Kim Kisslinger[∥], Hannah Clevenson[†], and Dirk Englund[†*]

[†] *Department of Electrical Engineering and Computer Science, and Research Lab of Electronics, Massachusetts Institute of Technology, 77 Massachusetts Ave., Building 36-575. Cambridge, MA 02139, USA*

[‡] *Department of Electrical Engineering, Columbia University, New York, NY 10027, USA*

[∥] *Center for Functional Nanomaterials, Brookhaven National Laboratory, Upton, NY 11973, USA*

KEYWORDS: implantation mask, nano-aperture, diamond color centers, Nitrogen Vacancy, spin chain, quantum computing.

## 1. Minimal line aperture of silicon hard mask

Patterns written upon ZEP resist on SOI by EBL and subsequent etched into the membrane on the reference sample are shown in Figure 1. A minimal width $W_{SI}$ = 25 nm was reached, but this was produced with significant degradation in the lines wall roughness.

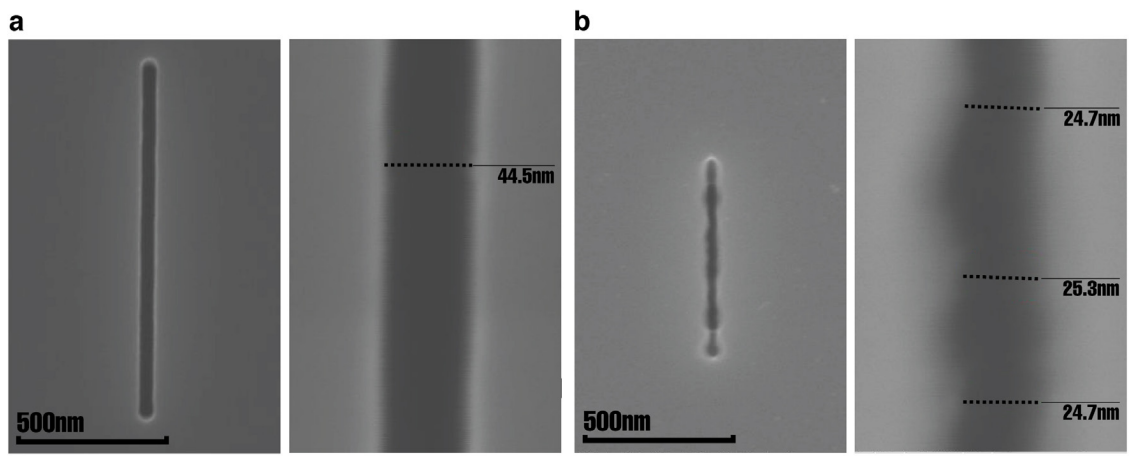

**Supplementary Figure 1. SEM images of line mask on the reference sample. a.** $W_{Si}$ = 45 nm line with a close up on the right. **b.** The minimal line width $W_{Si}$ = 25 nm with a close up of wall roughness on the right.

To achieve further decrease in pattern dimensions we have opted for two different modifications of this processing:

*1) Lithographic decrease in the patterned width:* This required thinning the resist (25% diluted ZEP) and SOI (160 nm silicon membrane). Despite significant charging-induced distortion, holes ~10nm wide were produced; however, the subsequent etching revealed that the aspect ratio of these holes is less than 2:1, i.e. these are too shallow to form a mask.

*2) Controlled vertical etching:* This technique is based on the inherent angularity of high aspect ratio dry etching. The lines were written on the undiluted ZEP with a nominal width $W_{Si} = 25$ nm and etched within a shorter time of 15 seconds, as compared to the 30 second long etching of Samples A and B, to allow for etching angularity. This resulted in ~10 nm wide line at the lower end of the mask (see Fig. 2b); however, the process exhibited low repeatability. This was attributed to the fact that the line profile is defined by the etching rate; thus, it is affected by the plasma stabilization period ("strike"). Since the etching during the strike cannot be controlled, arbitrary mask aperture profiles are produced.

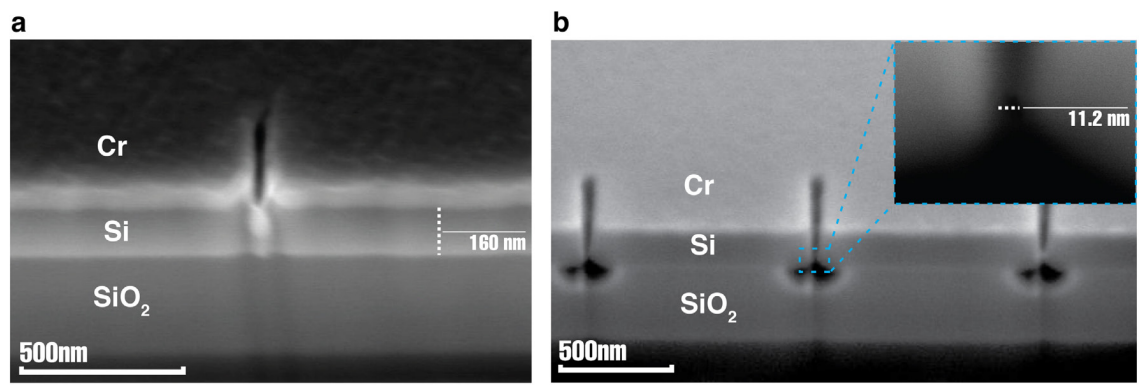

**Supplementary Figure 2. SEM images of implantation line scaling techniques a.** Decrease in the pattern: $W_{Si}=10$ nm line etched on a silicon membrane of 160nm and covered with Cr for improved imaging. **b.** Controlled vertical etching of 160 nm thick SOI. In the inset the angular aperture of the mask is depicted.

To conclude this discussion, optimization of the standard processing techniques was not successful in reducing the transferred pattern widths below 20 nm.

**2. Throughput Estimation**

Ideally, lines of $W_{Si} = 50$ nm and $L = 200$ nm written at a current of 250pA, energy of 100kV, and dosage of 600 µC/cm² are patterned at a nominal rate of $2.5\times10^5$ lines/min. Therefore, 2mm-size mask could be patterned in as low as 16min. However, in practice, this estimation does not account for several factors: 1) The high magnification mode of the JEOL JBX-6300FS which produces the highest resolution has a field of view (FOV) of 66 μm×66 μm. Therefore, switching from one FOV to another will result in a stage movement and stabilization periods. 2) To facilitate faster undercut of the mask in high concentration HF, every line, composed of a single FOV, is surrounded by an ~18 μm pattern of 250 nm square etching holes. These allow for undercut time of ~15-20 min. 3) Each mask is defined by 0.5 μm wide lines that surround its 8 mm perimeter. 4) Each mask includes line notation that can be read in optical microscope, to allow for identification of specific mask regions. 5) Finally, there are necessary EBL calibration times in between the mask writings. Therefore, in practice, patterning one 2 mm × 2 mm mask takes 9-10 hours of EBL.

## 3. NV localization

Figure 3a shows the lines of fluorescent NVs of L=200-400 nm and $W_{Si}$=25-35 nm produced in a reference sample. The number of NVs in a line depends on the aperture dimensions, implantation straggle (3-5nm), implantation dose ($8\times10^{12}$ ions/cm$^2$), and conversion yield (1-3%) from implanted nitrogen into NV. In Figure 3b-d lines occupied by 2-4 NVs are analyzed. A line containing four NVs is shown in Figure 3d, and a minimum distance between two near NVs was measured to be 8±4 nm. This distance is two times lower than the minimum pitch produced by AFM-tip[1]. Comparable (9 nm) distances between two NVs are produced in a 3-NV line (see Fig. 3c), while for the resolved 2 NV line the distance is 33 nm (see Fig. 3b). In general, the spread in the NV distance exemplified in Figure 3 originates from the probabilistic nature of NV formation. Obviously, this reduces the probability in producing large spin arrays; thus, stressing the importance of screening large number of NV lines to obtain a configuration useful for quantum information applications as is exhibited in the fabrication strategy for Sample B.

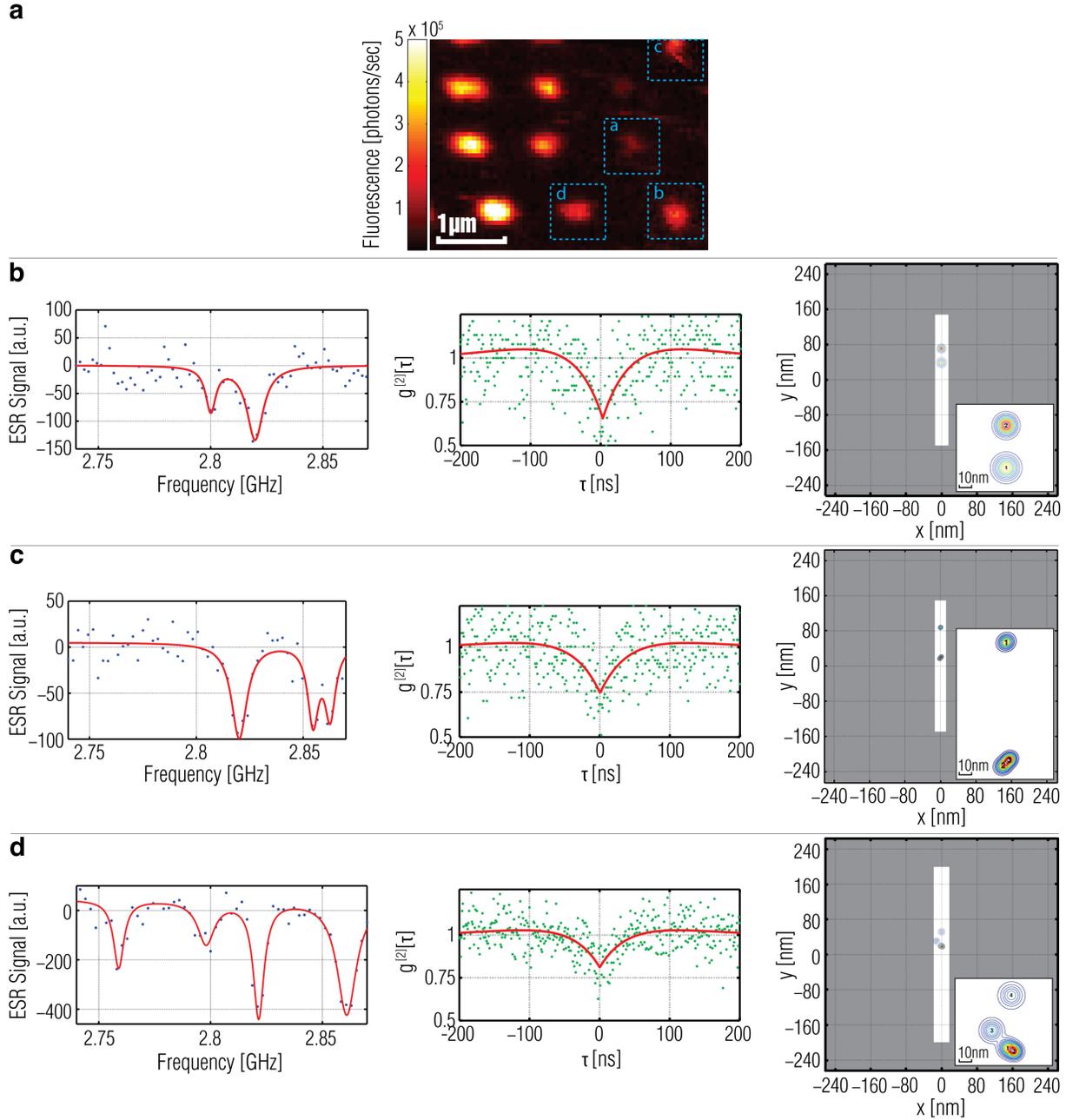

**Supplementary Figure 3. Analysis of several implantation lines of various dimensions on the reference sample. a,** Confocal scans of various length and width lines. The resolved lines are marked on the scan, where (a) marks a single NV line (L=200 nm, W=30 nm). **b-d.** Analysis

of individual lines presenting $g^{(2)}$ (top), ESR (middle), NV distribution based on 20 super-resolution measurements. The area outside the line is shaded. **b.** A line occupied by 2 NVs (L=300 nm and W=30 nm) described by ESR, $g^{(2)}$=0.65, and NV distribution with an NV-NV distance of 33±7 nm. **c.** A line occupied by 3 NVs (L=300 nm and W=25 nm) described by ESR, $g^{(2)}$=0.748, and NV distribution with NV-NV distances of d(NV2,NV3)= 9±4 nm, d(NV1,NV3)= 100±5 nm. **d.** A line occupied by 4 NVs (L=400 nm and W=35 nm): ESR indicating four NV orientations, $g^{(2)}$=0.81, spatial distribution of NVs with pitches of d(NV1,NV2)= 8±4 nm, d(NV1,NV3)= 19±6 nm, d(NV1,NV4)= 40±7 nm.

4. **Sample Characterization Set-Up**

The samples were optically characterized with a Zeiss Objective (NA = 1.3, EC Plan-NeoFluar 100x) on a commercial microscope (Zeiss Axio Observer.Z1m) outfitted with a ProEM-512K CCD. The 532nm excitation source (Coherent Verdi G5) was passed through an acoustic-optical modulator (Gooch & Housego 35085-0.5) as well as a quarter-wave plate to achieve uniform excitation of all NV orientations, and focused onto the back-aperture of the objective for wide-field illumination (~ 10 x 10 $\mu m^2$ FWHM). The resulting emission was filtered using 635nm long-pass (Semrock) and 532 nm notch (Thorlabs) filters.

Alternatively, for confocal excitation, the collimated beam was aligned directly into the objective, and fluorescence collected into a single-mode fiber (SMF) and measured by a single photon detector (Excelitas SPCM-AQ4C-IO). Auto-correlation measurements were done by coupling the SMF output to a fiber directional coupler (FC632-50B-FC), splitting the output between two single photons detectors whose correlation was measured by a time-correlated single photon counting module (Picoharp 300). For these measurements, a 650 nm long-pass filter (Thorlabs) was used in addition to a 532 nm notch (Thorlabs).

For both wide-field and confocal spin manipulation, microwave excitation was achieved through a 15-30 μm wire placed approximately 15-30 μm away from the region of interest. The microwave source (Rohde & Schwarz SMIQ03s) was directed through a TTL–controlled switch (Minicircuits ZASWA-2-50DR+) before being amplified (Minicircuits ZHL-16W-43+) and delivered to the sample. Timing of the pulses was achieved with a Pulse Blaster ESR Pro (SpinCore).

5. Sample A - NV Distribution characterization

For estimating the distribution of implanted NVs through this mask, the super-resolution technique as performed on the other samples was not successful; we consistently measured poor ODMR contrast due to a strained crystallographic structure from mechanical polishing of the surface, which was relieved in subsequent samples via oxygen RIE preparation[2]. Thus, it was critical to find a region with sparsely implanted nitrogen such that the average NV- in each aperture was close to unity. Such centers were confirmed to be single by photon auto-correlation. Then, by taking high SNR PL images, Gaussian fitting was used to estimate with high precision the center of the single NVs. Given an array of sparsely populated apertures, the position of the single NVs were projected onto the axis perpendicular to the apertures. By keeping track of the columns in which each NV was situated, a linear fit (with quadratic correction) was used to estimate the distribution of the single NVs about an unbiased estimation of the center of each aperture. The histogram shown in the main text is the residual distribution of single NVs about this unbiased estimator of the center. In other words, the displacement of all single NVs from the center is collapsed onto a single aperture to measure the implantation distribution.

## 6. Conformal ALD of circular apertures

In Fig. 4a conformal ALD of 22.5 nm was applied to circular apertures of 76 and 86 nm diameters in silicon. This resulted in shrinking aperture diameter towards 30 and 40 nm, respectively. Further shrinkage of more complex geometry is shown in Fig. 4b on the MIT logo. This initial results show that the processing approach described in the letter can be applied to circular and complex aperture geometries.

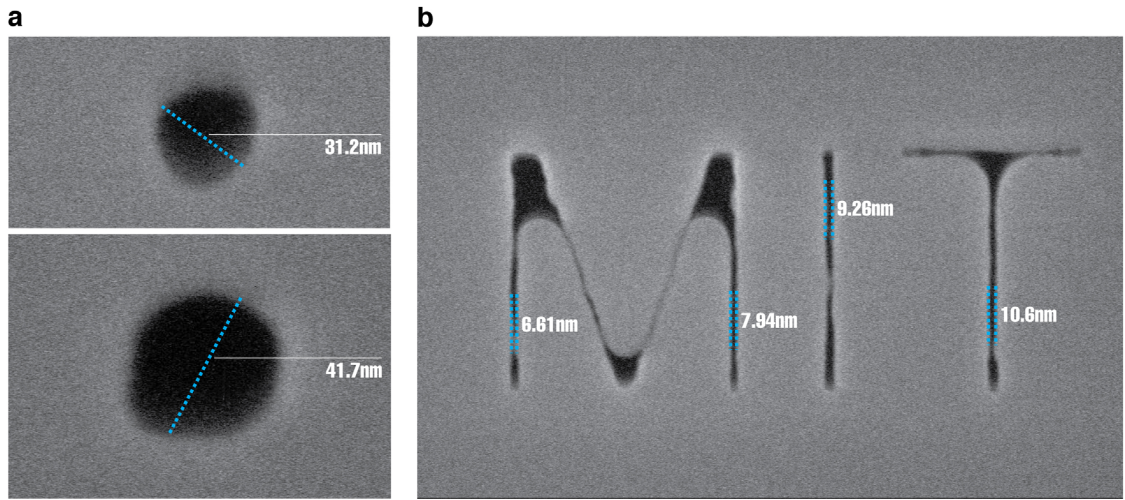

**Supplementary Figure 4. Conformal filling of curved structures at 270 nm thick SOI. a.** Two circles with initial diameter of 76 nm (top) and 86 nm (nm) covered by 22.5 nm of alumina. **b.** MIT logo with identical coverage.